\documentclass[aps,pra,twocolumn,showpacs,preprintnumbers,superscriptaddress,amsfont,graphicx,nofootinbib]{revtex4-1}

\usepackage[dvipdfm]{graphicx} 
\usepackage{bm}
\usepackage{amssymb}
\usepackage{amsmath}
\usepackage{color}
\usepackage{cancel}
\usepackage{comment}
\usepackage{here}
\usepackage{multirow}
\usepackage{url}
\usepackage{braket}
\usepackage{tikz}
\usepackage{float}
\usepackage{tikz-feynhand}
\usepackage{mathrsfs}
\usepackage{ulem}

\usepackage[toc,page]{appendix}

\allowdisplaybreaks

\begin{document} 

\title{On the criterion of perturbativity with the mass-dependent beta function \\ in extended Higgs models}

\author{Shinya Kanemura}
\email{kanemu@het.phys.sci.osaka-u.ac.jp }
\affiliation{Department of Physics, Osaka University, Toyonaka, Osaka 560-0043, Japan}

\author{Yushi Mura}
\email{y\_mura@het.phys.sci.osaka-u.ac.jp }
\affiliation{Department of Physics, Osaka University, Toyonaka, Osaka 560-0043, Japan}

\preprint{OU-HET-1207}

\begin{abstract}
In order to realize electroweak first order phase transition, a category of extended Higgs models with relatively large self-coupling constants is often considered.
In such a scenario, the running coupling constants can blow up at an energy scale much below the Planck scale.
To clarify the allowed parameter space of the model, it is important to evaluate the scale where perturbation calculation breaks down.
In the renormalization group equation analysis using the mass-independent renormalization scheme, we need the matching condition to connect the low energy theory and the high energy theory.
Consequently, the analysis depends on the detail of the matching condition.
On the other hand, the analysis with the mass-dependent beta function is performed in a simpler way, because the threshold effect is automatically included in the beta function.
Therefore, in this paper, using the mass-dependent beta function at the one-loop level, we discuss the application limit of perturbation calculation.
First, in a toy model with two complex scalar fields, we explain the essence of our method and compare our results with those based on the mass-independent beta function.
We then apply our method to the more realistic model; i.e., the inert doublet model.
We find that in our method, in which the threshold effect is automatically included, the energy scale where the perturbation calculation breaks down can be higher than the one using the mass-independent beta function, especially when the blow up scale is relatively low.
\end{abstract}

\maketitle

\section{Introduction}
In spite of dedicated experimental and theoretical efforts after the discovery of a Higgs boson in 2012, the structure of the Higgs sector and dynamics of electroweak (EW) symmetry breaking remain unknown.
Aspect of EW phase transition, which is fatally important for our deeper understanding of the thermal history of the Universe, is still mysterious.
Detailed understanding of the Higgs sector is also important to explore new physics beyond the SM such as baryon asymmetry of the Universe.

For example, in the scenario based on EW baryogenesis~\cite{Kuzmin:1985mm}, extended Higgs sectors are introduced to realize Strongly First Order EW Phase Transition (SFOPT), while it is smooth crossover in the SM~\cite{Kajantie:1996mn,DOnofrio:2015gop}.
Models for EW baryogenesis have been studied in the literature.
(For reviews, see Refs.~\cite{Cohen:1993nk,Trodden:1998ym,Riotto:1999yt,Morrissey:2012db}, and some of the recent developments are given in Refs.~\cite{Cline:2011mm,Dorsch:2016nrg,Cline:2021iff,Enomoto:2021dkl,Enomoto:2022rrl,Kanemura:2023juv}.)
The SFOPT is often caused by the non-decoupling quantum effect of extra scalar bosons in the effective potential.
Such an effect can also affect the coupling constants related to the Higgs boson such as the triple Higgs boson coupling~\cite{Kanemura:2002vm,Kanemura:2004mg, Kanemura:2004ch}.
Therefore, the SFOPT can be tested by measuring the triple Higgs boson coupling at future collider experiments.
Evidence of the SFOPT can also be found as characteristic stochastic gravitational waves, which may be observed at future space-based interferometers~\cite{Grojean:2006bp,Kakizaki:2015wua,Hashino:2016rvx,Hashino:2018wee}, and as primordial black holes with a specific mass~\cite{Hashino:2021qoq,Hashino:2022tcs}.

The ultraviolet (UV) behavior of the extended Higgs sector for the SFOPT, based on the analysis using Renormalization Group Equations (RGEs), can be drastically different from that of the SM.
In the SM, the effective Higgs self-coupling constant falls down to be negative at a scale between $10^{9}$ GeV and the Planck scale~\cite{Buttazzo:2013uya}.
It has turned out that the EW vacuum is meta stable~\cite{Bezrukov:2012sa,Degrassi:2012ry,Buttazzo:2013uya,Bednyakov:2015sca}.
On the other hand, in extended Higgs models with the non-decoupling property, due to quantum effects of extra scalar bosons, the effective coupling constants can blow up at a scale much below the Plank scale, above which the theory cannot be applied~\cite{Inoue:1982ej,Flores:1982pr,Kominis:1993zc,Nie:1998yn,Kanemura:1999xf,Ferreira:2009jb,Goudelis:2013uca,Chowdhury:2015yja,Kanemura:2017gbi,Kanemura:2019slf,Aiko:2023xui}.

Energy dependence of effective coupling constants is determined by the beta function.
The modified Minimal Subtraction ($\overline{\mathrm{MS}}$) scheme~\cite{tHooft:1972tcz}, which is one of the treatments for renormalization, provides the beta function as a function of only the coupling constants, and masses and the renormalization scale do not explicitly appear in the beta function.
This Mass Independent (MI) beta function has been widely used to study various high energy phenomena.

The coupling constants in the MI scheme are not directly connected to physical observables.
In addition, in the method using the MI beta function, no mechanism is equipped to treat the effect of appearance of a heavy particle around the scale of the mass.
Therefore, to calculate running coupling constants from a low energy effective theory to a theory where heavy particles become active, matching conditions for the coupling constants have to be taken into account.
This is so-called Threshold Correction (TC)~\cite{Weinberg:1980wa}, and the effects of the heavy particles are switched on as a step function in the beta function.
In many discussions, e.g. in the Grand Unified Theory (GUT)~\cite{Hall:1980kf,Ellis:1991ri, Ellis:1992kb, Yamada:1992kv, Hagiwara:1992ys}, it has been shown that TC of the heavy particles plays an important role.

For example, when we discuss the renormalization group analysis at the one loop level, as the tree level TC, the step function is often used to switch on the contribution of a massive particle at the mass scale of the particle.
However, it is not taken into account non-zero probability to find the particle in loop diagrams below that mass scale.
If we choose a different matching scale, a logarithmic difference arise~\cite{Braathen:2017jvs}.
Furthermore, additional error from higher-order loop corrections can enter into the matching conditions, when the MI coupling constants are expressed in terms of the physical observables as indicated in Refs.~\cite{Weinberg:1980wa} and \cite{Buttazzo:2013uya,Spencer-Smith:2014woa,Braathen:2017jvs}.
Therefore, much efforts has been elaborated to reduce this uncertainty by taking into account higher order TC.

On the other hand, in the the momentum subtraction (MOM) renormalization scheme, the coupling constants are directly connected to the physical observables, and the threshold effects are included in the counter term.
As a result, in the method using the Mass Dependent (MD) beta function, which is obtained by the MOM scheme, the quantum effects of heavy particles are naturally included.
Therefore, the MD beta functions are not suffered from the uncertainty of the TC.
This method has been used in the early studies~\cite{DeRujula:1976edq, Georgi:1976ve} to discuss the energy dependence of the effective QCD coupling constant for different number of the quarks in the theory.
In addition, EW running coupling constants~\cite{Ross:1978wt} and vacuum stability~\cite{Spencer-Smith:2014woa} in the SM, the proton decay~\cite{Goldman:1980ah} and unification of the coupling constants~\cite{Ross:1978wt,Faraggi:1993qb,Bagger:1995bw,Bastero-Gil:1995wxn,Binger:2003by} in the GUT, etc., also have been studied in this method.
However, it has not been applied to the physics of extended Higgs sectors yet.

In this paper, we discuss UV behaviors of effective coupling constants in extended Higgs sectors using the RGEs with the MD beta functions.
This paper is organized as follows.
In sec.~\ref{sec:2}, we first introduce the MD beta function in the MOM scheme.
In sec.~\ref{sec:3}, we then demonstrate to evaluate effective coupling constants in a toy model as a simple example.
In sec.~\ref{sec:4}, we apply our method to more realistic models; i.e., the Inert Doublet Model (IDM)~\cite{Deshpande:1977rw}.
We find that the upper bound on the model is higher than that obtained by the analysis of the MI beta function, especially when such a limit is near the TeV scale.
In sec.~\ref{sec:5}, we give some comments against the results, and then we present the conclusion.

\section{Mass dependent beta function\label{sec:2}}

\begin{figure}[t]
  \centering
  \includegraphics[width=1.\linewidth]{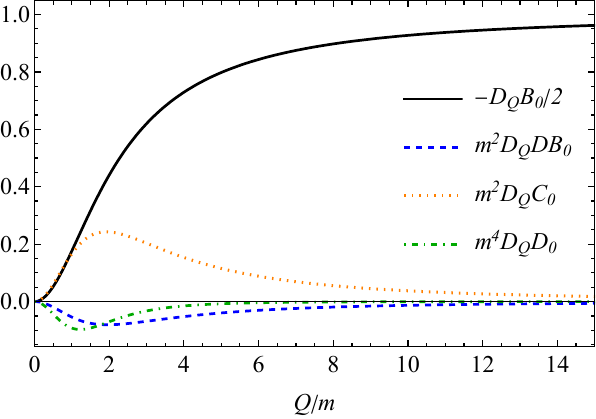}
  \caption{Derivatives with $\mathcal{D}_Q \equiv Q (\partial / \partial Q)_0$ of $-\mathrm{B}_0/2$ (black solid), $m^2 \mathrm{DB}_0(\mathrm{DB}_0 \equiv \partial \mathrm{B}_0/\partial k^2)$ (blue dashed), $m^2 \mathrm{C}_0$ (orange dotted) and $m^4 \mathrm{D}_0$ (green dot-dashed).
  Corresponding the particle to be decoupled from diagrams for $Q/m \to 0$, these functions become 0.
  For $Q/m \to \infty$, only the derivative of $\mathrm{B}_0$, which contains divergent parts, becomes unity.}
  \label{fig:LoopFunc}
\end{figure}

In order to introduce the MD beta function, we first see the $\lambda \phi^4$ scalar theory with the mass $m$.
The renormalization conditions for the $n$-point vertex function $\Gamma^{(n)}_{\phi \cdots \phi}(k_a, m^2, \lambda)$ are given by
\begin{align}
  &\Gamma^{(2)}_{\phi\phi}\Big|_{k^2 = m^2} = 0,~~ \frac{\partial}{\partial k^2 } \Gamma^{(2)}_{\phi \phi} \Big|_{k^2 = -Q^2} = 1, \notag \\
  &\Gamma^{(4)}_{\phi \phi \phi \phi} \Big|_{k_a\cdot k_b = -Q^2 \delta_{ab} + \frac{1}{3}Q^2 (1 - \delta_{ab})} = - \lambda,
  \label{eq:massdep_GL}
\end{align}
where $k_a$ ($a=1,\cdots,n$) are the external momenta.
The equations in the first line are the condition for the mass parameter $m$ to be the pole mass and the wave function to be unity at $k^2 = -Q^2$.
The equation in the second line requires that the four-point function coincides with the tree level value at the symmetric point where $k_a\cdot k_b = -Q^2 \delta_{ab} + \frac{1}{3}Q^2 (1 - \delta_{ab})$ are satisfied, and the UV divergence is renormalized in the MOM scheme.
From these renormalization conditions, at the one-loop level, we obtain the MD beta function as
\begin{align}
  \beta \Big(\lambda, \frac{Q}{m} \Big) &= \frac{3 \lambda^2}{16 \pi^2} \bigg[ -\frac{1}{2} \mathcal{D}_Q \mathrm{B}_0 \Big( -\frac{4}{3}Q^2, m^2, m^2 \Big) \bigg],
  \label{eq:massdep_betafunc}
\end{align}
where $\mathrm{B}_0$ is the Passarino--Veltman two-point scalar function~\cite{tHooft:1978jhc,Passarino:1978jh}, which contains an UV logarithmic divergence.
We define $\mathcal{D}_Q \equiv Q (\partial / \partial Q)_0$, where the subscript zero means derivative for the fixed bare parameters.
As a function of $Q/m$, $-\mathcal{D}_Q \mathrm{B}_0/2$ in Eq.~(\ref{eq:massdep_betafunc}) is shown in Fig.~\ref{fig:LoopFunc} as the black solid line.
This function takes a value between 0 and 1 for $0 < Q /m < \infty$ and represents how the particle with the mass $m$ contributes to the beta function at the scale $Q$.
If we take the limit $Q /m \to \infty$, Eq.~(\ref{eq:massdep_betafunc}) coincides with the well-known form given in the MI scheme (e.g. $\overline{\mathrm{MS}}$ scheme).
On the other hand, in the $\lambda \phi^4$ theory, the beta function becomes 0 for $Q/m \to 0$, because the quantum effect of the particle to the beta function decouples.
In Fig.~\ref{fig:LoopFunc}, we also show the lines corresponding to derivatives with $\mathcal{D}_Q$ of $m^2 \mathrm{DB}_0(\mathrm{DB}_0 \equiv \partial \mathrm{B}_0/\partial k^2)$ (blue dashed), $m^2 \mathrm{C}_0$ (orange dotted) and $m^4 \mathrm{D}_0$ (green dot-dashed), where $\mathrm{C}_0$ and $\mathrm{D}_0$ are the Passarino--Veltman three and four point scalar functions, respectively.
They are discussed later.

In the following, we apply the MD beta function to extended Higgs models.
We are interested in the case where additional scalar bosons obtain masses mainly from the EW symmetry breaking.
Such a case is motivated for a SFOPT~\cite{Funakubo:1993jg,Davies:1994id,Cline:1996mga}.
We then show importance of the method using the MD beta function, in which the decoupling mechanism of heavy particles is naturally equipped.

Before moving the discussion, we here comment on the renormalization scheme independence.
Although the explicit form of the vertex functions is different among the renormalization schemes, physical observables, of course, are scheme independent.
However, due to the truncation at a finite order of perturbation calculation, different results can appear depending on the schemes.
For example, it has been known that the scale dependence of the QCD coupling constant at the two-loop level is different between $\overline{\mathrm{MS}}$ and MOM schemes~\cite{Jegerlehner:1998zg}.

\section{Simplest example of the non-decoupling case\label{sec:3}}
We here discuss a toy model which consists of two complex scalar fields $\phi_1$ and $\phi_2$ to see the idea of our method.
We impose two global symmetries $U(1)_1$ and $U(1)_2$ under the transformations $\phi_1 \to \phi_1 e^{i\theta_1}$ and $\phi_2 \to \phi_2 e^{i\theta_2}$, respectively.
The renormalizable Lagrangian is given by
\begin{align}
  \mathcal{L} &= \sum_{i=1}^2 \Big( |\partial_\mu \phi_{i}^b|^2 - \mu_{i,b}^2 |\phi_i^b|^2 \Big) \notag \\
  &-\frac{1}{2}\lambda_1^b |\phi_1^b|^4 -\frac{1}{2}\lambda_2^b |\phi_2^b|^4 - \lambda_3^b |\phi_1^b|^2|\phi_2^b|^2,
  \label{eq:toy_lagrangian}
\end{align}
where $\mu_i^2~(i=1,2)$ and $\lambda_j ~(j=1,2,3)$ are real parameters, and the subscript $b$ represents the bare parameter.
We assume that $U(1)_1$ is spontaneously broken while $U(1)_2$ is not.

At the tree level, the subscript $b$ can be dropped.
We parametrize the field $\phi_1$ as $\phi_1 = (v + \rho + i \eta) / \sqrt{2}$, where $v$ is the Vacuum Expectation Value (VEV).
The stationary condition gives the relation among the parameters: 
\begin{align}
  \frac{\partial V}{\partial \rho} \bigg|_{\{\rho,\eta,\phi_2\}=\bm{0}} = 0 ~ \Leftrightarrow ~ v^2 = \frac{-2 \mu_1^2}{\lambda_1}.
  \label{eq:toy_station}
\end{align}
Masses of the physical states $\rho$, $\eta$ and $\phi_2$ are given by $m_{\rho}^2 = \lambda_1 v^2$, $m_{\eta}^2 = 0$ and $m_{\phi_2}^2 = \mu_2^2 + \frac{1}{2}\lambda_3 v^2$.
The field $\eta$ is the Nambu--Goldstone Boson (NGB) related to the spontaneously breaking of $U(1)_1$.

In the case of $\lambda_1 v^2 \simeq \lambda_3 v^2 \ll \mu_2^2$, we have $m_{\phi_2}^2 \simeq \mu_2^2$.
The low energy effective theory, whose cut off scale corresponds to $|\mu_2|$, is described only by the fields $\rho$ and $\eta$, while the effect of $\phi_2$ decouples by $1/\mu_2^2$ due to the decoupling theorem~\cite{Appelquist:1974tg}.
On the other hand, for $\mu_2^2 \lesssim \lambda_3 v^2$, the effect of $\phi_2$ cannot decouple in the low energy effective theory.
Non-decoupling quantum effects of $\phi_2$ can appear as powerlike or logarithmic mass contributions~\cite{Kanemura:2004mg}.

In this model, non-decoupling property is determined by the size of $\lambda_3$; i.e., if the mass of the additional scalar boson $m_{\phi_2}$ is mainly given by $\lambda_3 v^2$, quantum corrections to the low energy observables are large.
In such a case, however, the quartic coupling constants can quickly blow up.
In other words, if we impose that the coupling constants do not exceed a critical value (such as $4 \pi$) below a cut off scale $\Lambda$, the coupling constants are constrained from above as a function of $\Lambda$.
Such a theoretical bound is called as the triviality bound~\cite{Dashen:1983ts}.

In the following, we study renormalization group analysis with the MD beta function, and give a constraint from the triviality bound at the one-loop level.
We define seven renormalization constants as
\begin{align}
  &\phi_i = Z_{\phi_i}^{-\frac{1}{2}} \phi_i^b, ~~\mu_i^2 = Z_{\mu_i}^{-1} Z_{\phi_i} \mu_{i,b}^2, ~~\lambda_1 = Z_{\lambda_1}^{-1} Z_{\phi_1}^2 \lambda_1^b,\notag \\
  &\lambda_2 = Z_{\lambda_2}^{-1} Z_{\phi_2}^2 \lambda_2^b, ~~ \lambda_3 = Z_{\lambda_3}^{-1} Z_{\phi_1}Z_{\phi_2} \lambda_3^b.
  \label{eq:toy_bareandren}
\end{align}
They are determined by seven renormalization conditions, which are discussed in the following paragraphs.

First, the renormalized VEV is defined as the right-hand side of Eq.~(\ref{eq:toy_station}) with the renormalized coupling constants, and we define the renormalized shifted fields as $\phi_1 = (v + \rho + i\eta) /\sqrt{2}$.
The tadpole term for the field $\rho$, which is proportional to $Z_{\mu_1} - Z_{\lambda_1}$, appears in the Lagrangian.
We impose a renormalization condition that the left-hand side of Eq.~(\ref{eq:toy_station}) is satisfied at the one-loop level.
The tadpole term is eliminated by this condition, and then $m_\eta$ is zero at the one-loop level.

Second, we set the renormalization conditions as
\begin{align}
  &\Gamma^{(2)}_{\rho \rho}\Big|_{k^2 = m_{\rho}^2} = \Gamma^{(2)}_{\phi_2 \phi^\dagger_2}\Big|_{k^2 = m_{\phi_2}^2} = 0, \notag \\
  &\frac{\partial}{\partial k^2} \Gamma^{(2)}_{\rho \rho} \Big|_{k^2 = -Q^2} = \frac{\partial}{\partial k^2} \Gamma^{(2)}_{\phi_2 \phi^\dagger_2} \Big|_{k^2 = -Q^2} = 1,
  \label{eq:toy_rencon1}
\end{align}
where the first and second lines are the pole mass conditions and the wave function renormalization conditions, respectively.
From the pole mass conditions, we obtain $\mathcal{D}_Q m_\rho^2 = \mathcal{D}_Q m_{\phi_2}^2 = 0$.

Finally, we impose two conditions for $Z_{\lambda_3}$ and $Z_{\lambda_2}$ as
\begin{align}
  \Gamma^{(4)}_{\rho \rho \phi_2 \phi^\dagger_2} \Big|_{k=\mathrm{sym.}} = -\lambda_3, ~~~ \Gamma^{(4)}_{\phi_2 \phi_2^\dagger \phi_2 \phi_2^\dagger} \Big|_{k=\mathrm{sym.}} = -2 \lambda_2,
  \label{eq:toy_rencon2}
\end{align}
where the renormalization point is the symmetric point.
At this point, the four-point functions coincide with the tree-level values.
In appendix~\ref{sec:app}, explicit formulae for these vertex functions are shown at the one-loop level.

With the renormalization conditions imposed above, we derive the RGE for the renormalized effective action $\Gamma(\phi_i;m_{\rho},m_{\phi_2},\lambda_j;Q)$.
Here we have changed independent parameters from $(\mu_1^2,\mu_2^2)$ to $(m_{\rho}^2,m_{\phi_2}^2)$.
Notice that $m_{\rho}$ and $m_{\phi_2}$ are scale invariant, and we obtain 
\begin{align}
  \bigg( Q \frac{\partial}{\partial Q} - \gamma_{\phi_i} \phi_i \frac{\delta}{\delta \phi_i} + \beta_{\lambda_j} \frac{\partial}{\partial \lambda_j} \bigg) \Gamma = 0,
\end{align}
where $\gamma_{\phi_i} \equiv - \mathcal{D}_Q \log \phi_i$ and $\beta_{\lambda_j} \equiv \mathcal{D}_Q \lambda_j$.

At the one-loop level, it can be shown that
\begin{align}
  &\beta_{\lambda_3} = \frac{1}{16 \pi^2} \bigg\{ \lambda_1 \lambda_3 + 3 \lambda_1 \lambda_3 f^Q_{m_{\rho}, m_{\rho}} +4 \lambda_3^2 f^Q_{m_{\phi_2}, m_{\rho}} \notag \\
  &+4 \lambda_2 \lambda_3 f^Q_{m_{\phi_2}, m_{\phi_2}} +\mathcal{D}_Q (\mathrm{DB}_0, \mathrm{C}_0, \mathrm{D}_0~\mathrm{terms}) \bigg\}.
  \label{eq:toy_mdbeta}
\end{align}
Similarly, $\beta_{\lambda_1}$ and $\beta_{\lambda_2}$ can also be obtained.
We have denoted the brackets in the right hand side of Eq.~(\ref{eq:massdep_betafunc}) as $f^Q_{m,m}$.
As shown in Fig.~\ref{fig:LoopFunc}, only the derivative of $\mathrm{B}_0$, namely $f$, is relevant to the beta functions in the limit $Q/m \to \infty$, while the other functions $\mathcal{D}_Q \mathrm{DB}_0, \mathcal{D}_Q \mathrm{C}_0$ and $\mathcal{D}_Q \mathrm{D}_0$ are not.
Although the beta functions $\beta_{\lambda_2}$ and $\beta_{\lambda_3}$ depend on $Q/m$, they coincide with the MI beta functions for the high energy limit, because $f \to 1$ for $Q/m \to \infty$. 
In the limit $Q/m \to 0$, the contributions from the massive particles decouple, and only the UV divergent diagrams, in which the NGBs are involved, contribute to $\beta_{\lambda_2}$ and $\beta_{\lambda_3}$.
In the numerical analysis below, we include the contributions from $\mathrm{DB}_0, \mathrm{C}_0$ and $\mathrm{D}_0$ in the MD beta functions.

Since we have imposed the tadpole condition in Eq.~(\ref{eq:toy_station}) and the on-shell condition for $m_\rho$ in Eq.~(\ref{eq:toy_rencon1}), $\beta_{\lambda_1}$ is asymptotically zero for $Q/m \to \infty$ in this model.
On the other hand, $\Gamma_{\rho\rho\rho\rho}^{(4)}$ depends on the external momenta, and corrections of $O(\frac{\lambda}{16\pi^2} \log s/m_{\rho}^2)$  arise at the one-loop level, where $\sqrt{s}$ is the center-of-mass energy. 
In the following, we focus on the energy dependence of $\lambda_j (j=1,2,3)$.

In order to obtain analytic formulae for the vertex functions, we use \texttt{FeynCalc}~\cite{Mertig:1990an, Shtabovenko:2016sxi, Shtabovenko:2020gxv}, \texttt{FeynArts}~\cite{Kublbeck:1990xc, Hahn:2000kx} and \texttt{FeynRules}~\cite{Christensen:2008py, Alloul:2013bka}.
For numerical value of the Passarino--Veltmann functions, we use \texttt{LoopTools}~\cite{Hahn:2000jm} and \texttt{FeynHelpers} package~\cite{Shtabovenko:2016whf}.

\begin{figure}[t]
  \centering
  \includegraphics[width=1.\linewidth]{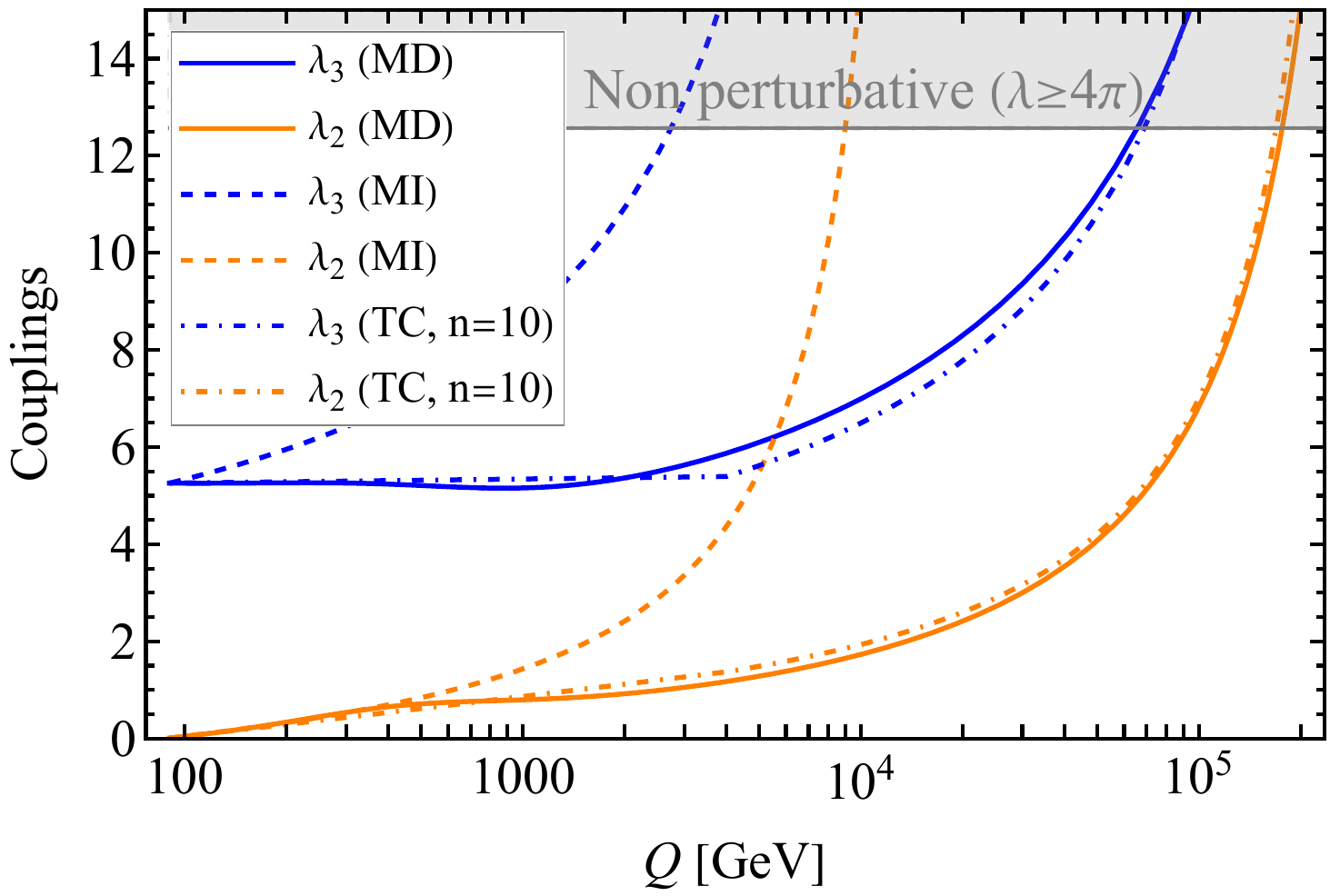}
  \caption{
  Energy dependence of $\lambda_3$ (blue) and $\lambda_2$ (orange) with the MD (solid), MI (dashed) and TC (dot-dashed) beta functions in the $U(1)_1 \times U(1)_2$ model.
  With the MD beta functions, which include a decoupling mechanism of the heavy particles by the factor $f$, the scale $\Lambda_{4 \pi}$ is shifted to the high energy region compared to the one with the MI beta functions.}
  \label{fig:Running}
\end{figure}
In Fig.~\ref{fig:Running}, the scale dependence of $\lambda_3$ and $\lambda_2$ with the MD beta functions are shown as the blue and orange solid lines, respectively.
We take the input parameters as $(v, m_{\rho}, m_{\phi_2})=(246,125, 400)$ GeV.
We consider the non-decoupling case with large $\lambda_3$, and we set $\lambda_3(Q_0) = 5.3$ and $\lambda_2(Q_0) = 0.01$ as input values at $Q = Q_0 \equiv 90$ GeV.
The dashed lines are the solutions of the MI beta functions obtained by the $\overline{\mathrm{MS}}$ scheme.

We define the scale $\Lambda_{4\pi}$ at which the largest coupling constant in the model becomes $4 \pi$.
As we can see from Fig.~\ref{fig:Running}, $\Lambda_{4\pi}$ is around $3$ TeV for the results with the MI beta functions.
However, $\Lambda_{4\pi} \simeq 60$ TeV for the results with the MD beta functions.
This is because the MD beta functions include the reduction factor $f$ for the heavy particles.
Consequently, running of the coupling constants delays.
Even if $\lambda_2$ is taken to be large at $Q=Q_0$, this running behavior is unchanged.
As we see in the next section, the scheme difference in the running coupling constants is prominent for the non-decoupling case.

In Fig.~\ref{fig:Running}, we also show the case with the improved MI beta functions, in which the TC is added by hand, as the dot-dashed lines.
As the TC, we define step functions $\theta^n (Q/m_{\phi_2})$ which take 1 (0) for $Q \ge n m_{\phi_2}$ $(Q < n m_{\phi_2})$.
The improved MI beta functions of $\lambda_j$ are given by
  \begin{align}
  &\beta_{\lambda_1}^{\mathrm{TC}} = \frac{1}{16 \pi^2} \big( 10 \lambda_1^2 + 2 \lambda_3^2 \theta^n \big), \notag \\
  &\beta_{\lambda_2}^{\mathrm{TC}} = \frac{1}{16 \pi^2} \big( 10 \lambda_2^2 \theta^n + 2 \lambda_3^2 \big), \notag \\
  &\beta_{\lambda_3}^{\mathrm{TC}} = \frac{1}{16 \pi^2} \Big( 4 \lambda_1 \lambda_3 + 4 (\lambda_2 \lambda_3 + \lambda_3^2 )\theta^n \Big).
  \label{eq:toy_thbeta}
\end{align}
In Fig.~\ref{fig:Running}, the solutions of the MI beta functions with the TC for $n=10$ are shown as the dot-dashed lines.
Corresponding to $f\simeq 1$ at $Q/m \simeq 10$ as shown in Fig.~\ref{fig:LoopFunc}, the solutions of the MI beta functions with the TC for $n=10$ behave similarly to those with the MD beta functions.

We note that, however, the treatment of the TC with one matching scale contains uncertainty where we insert the threshold step function~\cite{Braathen:2017jvs}.
Therefore, for more realistic models in which multiple heavy particles with different masses are introduced, it is rather complicated.
On the other hand, in our scheme with the MD beta function the treatment would be simpler for such a case.
\begin{figure*}[t]
  \centering
  \includegraphics[width=0.9\linewidth]{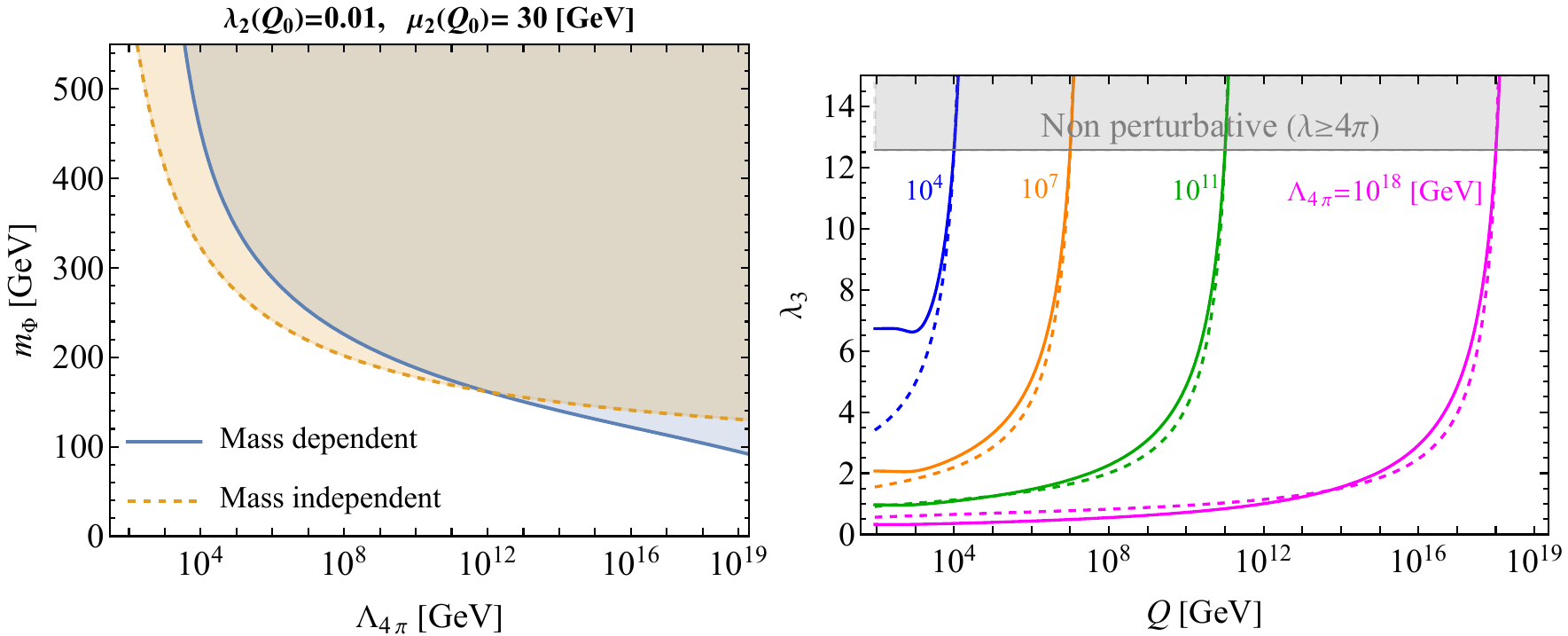}
  \caption{Left: Triviality bound for the degenerated masses of the additional scalar bosons~$m_{\Phi}$ in the IDM as a function of $\Lambda_{4 \pi}$.
  The blue solid and orange dashed lines show the results with the MD and MI beta functions, respectively.
  Right: The energy dependence of $\lambda_3$ for each fixed value of $\Lambda_{4 \pi}$; i.e. $10^{4}$ (blue), $10^{7}$ (orange), $10^{11}$ (green) and $10^{18}$ GeV (magenta).
  The solid and dashed lines are obtained by using the MD and MI beta functions, respectively.
  Due to the delay of running with the reduction factor $f$ in the MD beta function, the scheme difference stands out for the small cutoff scale, e.g. $\Lambda_{4\pi} = 10^4$ GeV.}
  \label{fig:SchemeDependence}
\end{figure*}

\section{Application to the extended Higgs models\label{sec:4}}
We here consider the IDM~\cite{Deshpande:1977rw}, in which an additional $SU(2)$ scalar doublet $\Phi_2$ is added to the SM, which is odd under the unbroken $Z_2$ symmetry.
The potential is given by 
\begin{align}
  &V = \mu_1^2 |\Phi_1|^2 +\mu_2^2 |\Phi_2|^2 + \frac{1}{2}\lambda_1 |\Phi_1|^4 + \frac{1}{2}\lambda_2 |\Phi_2|^4 \notag \\
  &+\lambda_3 |\Phi_1|^2 |\Phi_2|^2 +\lambda_4 |\Phi_2^\dagger \Phi_1|^2 +\lambda_5 \mathrm{Re}[(\Phi_2^\dagger \Phi_1)^2],
\end{align}
where $\Phi_1$ corresponds to the SM-like Higgs doublet.
The coupling constants $\lambda_1, \cdots, \lambda_5$ are all real.
Taking $\mu_1^2 < 0$ and $\mu_2^2>0$, the EW symmetry is spontaneously broken.

We parametrize the fields as $\Phi_1 = \big(G^+, (v + h + iG^0)/\sqrt{2} \big)^T$, $\Phi_2 = \big(H^+, (H + iA)/\sqrt{2} \big)^T$, where $v$ ($= \sqrt{-2 \mu_1^2 / \lambda_1}$) is the VEV, $h$ is the SM-like Higgs boson, and $H$ ($A$) and $H^\pm$ are the $Z_2$-odd CP-even, (CP-odd) and charged scalar bosons, respectively.
$G^+$ and $G^0$ are the NGBs absorbed as the longitudinal modes of the gauge bosons.
Squared masses are given by $m_h^2 = \lambda_1 v^2$, $m_{H^\pm}^2 = \mu_2^2 + \frac{1}{2}\lambda_3 v^2$ and $m_{H,A}^2 = m_{H^\pm}^2 + \frac{1}{2}(\lambda_4  \pm \lambda_5)v^2$ with the sign $+(-)$ for $H(A)$.

The discussion for the MD beta function is same as the case of the toy model except for the following points:
(i) $\lambda_4$ and $\lambda_5$ cause the mass difference among the additional neutral scalar bosons.
By performing the renormalization of $m_{H}$ and $m_{A}$ as the pole masses, $\mathcal{D}_Q(\lambda_4 v^2) = \mathcal{D}_Q(\lambda_5 v^2) = 0$ are shown.
As a result, $\beta_{\lambda_4}/\lambda_4$ and $\beta_{\lambda_5}/\lambda_5$ are equal to $\beta_{\lambda_1} / \lambda_1$.
(ii) Masses of the W and Z bosons and the top quark $t$ are renormalized on mass-shell, so that beta functions of the gauge coupling constants ($g_1$ and $g_2$) and top-Yukawa coupling constants ($y_t$) are related to that of $\lambda_1$; i.e., $2\beta_{g_1} /g_1 = 2\beta_{g_2} / g_2 = 2\beta_{y_t} / y_t = \beta_{\lambda_1} / \lambda_1$.
(iii) The Slavnov--Taylor identity~\cite{Slavnov:1972fg, Taylor:1971ff} among the renormalization constants does not hold in the MOM scheme.
This difficulty can be avoided~\cite{Rebhan:1985yf} by using the back ground field method~\cite{DeWitt:1967ub, DeWitt:1967uc, Abbott:1980hw,Denner:1994xt}.
We here choose the Feynman gauge for the internal propagators in the vertex functions.
We construct the model files for \texttt{FeynRules} for the IDM with the back ground field method based on Ref.~\cite{Degrande:2014vpa}.

In the IDM, one-loop effects of additional scalar bosons give positive contributions to the beta functions of the scalar coupling constants. 
If they are relatively large at the EW scale, the Landau pole can appear below the Planck scale.
Then, $m_H$, $m_A$ and $m_{H^\pm}$ are constrained by the triviality bound~\cite{Inoue:1982ej,Flores:1982pr,Kominis:1993zc,Nie:1998yn,Kanemura:1999xf,Ferreira:2009jb,Goudelis:2013uca,Chowdhury:2015yja,Kanemura:2017gbi,Kanemura:2019slf,Aiko:2023xui}.

In the left panel of Fig.~\ref{fig:SchemeDependence}, the triviality bound are shown as a function of $\Lambda_{4\pi}$ for the degenerated masses ($m_\Phi \equiv m_H=m_A=m_{H^\pm}$).
The solid and dashed lines are obtained by the MD and MI beta functions, respectively.
We take $\lambda_2(Q_0)=0.01$ and $\mu_2(Q_0) = 30$~GeV as an input parameter, so that the non-decoupling situation is realized by a large $m_{\Phi}$.
As shown in Fig.~\ref{fig:SchemeDependence}, for both the MI and MD beta functions, a larger $m_{\Phi}$ predicts a lower $\Lambda_{4\pi}$.
Remarkably, the scheme difference of the upper bound for fixed $\Lambda_{4\pi}$ is significant especially in such a non-decoupling case.

In the right panel of Fig.~\ref{fig:SchemeDependence}, the energy dependence of $\lambda_3$ are shown for each fixed cutoff scale $\Lambda_{4 \pi} = 10^{4}$ (blue), $10^{7}$ (orange), $10^{11}$ (green) and $10^{18}$ GeV (magenta).
As we can see from the blue lines in this figure, when $\lambda_3 (Q_0)$ is large with the fixed relatively small value of $\mu_2(Q_0)$ ($=30$ GeV), the scalar coupling constants quickly blow up for the MI beta function.
However, because the MD beta functions is relatively small in low energy regions due to the reduction factor $f$, the weaker triviality bound is predicted for the lower $\Lambda_{4\pi}$.
This behavior of the running coupling of $\lambda_3$ with the MD beta functions is drastically different from that of the running coupling with the MI beta functions.
The difference becomes more conspicuous in the case of the lower $\Lambda_{4\pi}$.

We note that in the IDM the similar non-decoupling effect can also be realized when a large mass difference appears among the additional scalar bosons.
Such large mass difference affects $\beta_{\lambda_3}$ and $\beta_{\lambda_2}$.
It makes these coupling constants rapidly blow up.
Even in this case, $\Lambda_{4\pi}$ calculated by the MD beta functions is higher than the one by the MI beta functions.
If a mass difference among the additional scalar bosons is large enough so that $\lambda_4$ or $\lambda_5$ is nearly $4 \pi$ at the renormalized point $Q_0$,
 quantum corrections to the related vertex functions become large, so that perturbation calculation breaks down as external momenta grow.

\section{Discussions and conclusions\label{sec:5}}
We here give comments against the results obtained above.

First, as shown in Figs.~\ref{fig:Running} and \ref{fig:SchemeDependence}, for the triviality bound the difference between the methods based on the MI and MD beta functions becomes large in the non-decoupling situation.
This is because the threshold effects are automatically included in the MD beta function.
The importance of the delayed running of the coupling constants can be understood as follows.
The MI beta functions of the scalar fields with the coupling $\lambda$ can be written as $\beta_{\lambda} \simeq c \lambda^2$ with a positive constant $c$.
On the other hand, for the analysis using the MD beta function with the automatically included delay of the running, the initial value of the MD coupling which deduces the same $\Lambda_{4 \pi}$ as the MI scheme satisfies the following relation,
\begin{align}
  \lambda(Q_0)^{\mathrm{MD}} - \lambda(Q_0) = \frac{c \lambda(Q_0)^2 \log \frac{Q_1}{Q_0}}{1 - c \lambda(Q_0) \log \frac{Q_1}{Q_0}},
\end{align}
where $Q_0$ is the renormalization scale, and $Q_1(>Q_0)$ is the scale where the running begins in the analysis based on the MD beta function.
The left-hand side describes the difference in the triviality bound between the renormalization schemes.
The right-hand side is the monotonic increasing function for the initial value $\lambda(Q_0)$ within the perturbative region.
Therefore, for the case of the non-decoupling situation where $\lambda(Q_0)$ is relatively large, the difference in the triviality bound becomes large due to the delayed running in the MD method.
In this scheme, such delayed running is automatically included, because the threshold effect appears in the counter term.

We emphasize that the both methods are in principle equivalent in the end.
In the case with multiple new particles exist the analysis with the MD beta function would be useful, because complicated treatment of TCs can be avoided. 

In this paper, the MD beta functions are evaluated at the one-loop level, and our method can also be applied to higher loop corrections.
By the analysis with the MI beta function at the two-loop level, it has been known that $\Lambda_{4\pi}$ can be higher than that of the one-loop level in the two Higgs doublet model~\cite{Chowdhury:2015yja}.
Traditionally, for the two-loop MI beta function, the one-loop level TC is used~\cite{Weinberg:1980wa}.
Recently, however, it has been indicated by J. Braathen et al.~\cite{Braathen:2017jvs} that in the non-decoupling case a $n$-loop level TC should be used for a $n$-loop level beta function.
It is known that $\Lambda_{4\pi}$ is larger by taking into account a higher order TC, because it tends to make the MI parameters which correspond to physics quantities smaller.
Therefore, it is expected that predictions in the MI scheme by using the one-loop beta function with the one-loop TC approach to the results in the one-loop MD method.

For the phenomenological purpose to evaluate the upper bound on the model for non-decoupling physics, e.g. EW baryogenesis, our one-loop analysis gives a very important insight.
The difference between both MD and MI schemes are shown to be large at the one-loop level.
As a result, the previous argument with the MI beta function at the one-loop level~\cite{Cline:2011mm,Dorsch:2016nrg}, in which it is shown that models of EWBG based on the two Higgs doublet extension are strongly constrained, would be modified.
This is the result of our physical MD beta function, in which the threshold effects are naturally included.

In this paper, UV behaviors of effective coupling constants have been discussed by using RGEs with the MD beta functions.
We have demonstrated to evaluate effective coupling constants in the $U(1)_1 \times U(1)_2$ model and the IDM.
We have found that $\Lambda_{4\pi}$ evaluated with the MD beta functions is higher than that obtained with the MI beta functions for the case when the non-decoupling effect is important.
We finally emphasize that the MD method discussed here can be applied to any new physics model with additional scalar fields.

\section*{Acknowledgements}
The work of S.~K. was supported by the JSPS KAKENHI Grant No.~20H00160 and No.~23K17691.
The work of Y.~M. was supported by the JSPS KAKENHI Grant No.~23KJ1460.

\appendix

\section{Formulae for the vertex functions at the one-loop level\label{sec:app}}

In this appendix, analytic formulae for the vertex functions in the toy model which have been introduced in sec.~\ref{sec:3} are given at the one-loop level.

The one-point vertex function of $\rho$ is given by
\begin{align}
  &\Gamma_{\rho}^{(1)}(0) = \frac{1}{16\pi^2} \bigg( \frac{3}{2} \lambda_1 v \mathrm{A_0}(m_{\rho}^2) + \lambda_3 v \mathrm{A_0} (m_{\phi_2}^2) \bigg) \notag \\
  &+ \frac{1}{2}v m_{\rho}^2 \big( Z_{\mu_1} - Z_{\lambda_1} \big),
\end{align}
where $\mathrm{A_0}$ is the Passarino--Veltman one-point scalar function~\cite{tHooft:1978jhc,Passarino:1978jh}, which does not have any momentum dependence.

The self energies for $\rho$ and $\phi_2$ fields are given by
\begin{align}
  &\Pi_{\rho\rho}(k^2) = 
  \frac{1}{16\pi^2} \bigg( \frac{3}{2} \lambda_1 \mathrm{A_0} (m_{\rho}^2) +\lambda_3 \mathrm{A_0}(m_{\phi_2}^2) \notag \\
  &+\frac{9}{2} \lambda_1 m_{\rho}^2 \mathrm{B_0}(k^2, m_{\rho}^2, m_{\rho}^2) +\frac{1}{2} \lambda_1 m_{\rho}^2 \mathrm{B_0}(k^2, 0,0) \notag \\
  &+ \lambda_3^2 v^2 \mathrm{B_0}(k^2, m_{\phi_2}^2, m_{\phi_2}^2) \bigg) +(Z_{\phi_1} -1) k^2 \notag \\
  &- \frac{1}{2} \Big( 3 (Z_{\lambda_1}-1) - (Z_{\mu_1}-1) \Big) m_{\rho}^2,
\end{align}
and
\begin{align}
  &\Pi_{\phi_2\phi_2^\dagger}(k^2) = 
  \frac{1}{16\pi^2} \bigg( 2 \lambda_2 \mathrm{A_0} (m_{\phi_2}^2) + \frac{1}{2}\lambda_3 \mathrm{A_0} (m_{\rho}^2) \notag \\
  &+ \lambda_3^2 v^2 \mathrm{B_0}(k^2, m_{\rho}^2, m_{\phi_2}^2 )  \bigg) +(Z_{\phi_2} - 1) k^2 \notag \\
  &- \Big( (Z_{\mu_2} - 1) \mu_2^2 + \frac{1}{2} (Z_{\lambda_3} - 1) \lambda_3 v^2 \Big).
\end{align}
The corresponding two-point vertex function is obtained by
\begin{align}
  \Gamma^{(2)}(k^2) = k^2 - m^2 + \Pi(k^2).
\end{align}

The four-point vertex functions at the symmetric point are given by
\begin{align}
  &16 \pi^2 \Gamma^{(4)}_{\rho \rho \phi_2 \phi^\dagger_2} \Big|_{k=\mathrm{sym.}} = - 16\pi^2 Z_{\lambda_3} \lambda_3 +\frac{3}{2} \lambda_1 \lambda_3 \mathrm{B_0} [ m_{\rho}^2, m_\rho^2 ] \notag \\ 
  &+ \frac{1}{2} \lambda_1 \lambda_3 \mathrm{B_0} [0,0] +2 \lambda_2 \lambda_3 \mathrm{B_0} [m_{\phi_2}^2, m_{\phi_2}^2 ] +2 \lambda_3^2 \mathrm{B_0} [m_{\phi_2}^2, m_\rho^2 ] \notag \\
  &+9 \lambda_1 \lambda_3 m_\rho^2 \mathrm{C_0} [m_\rho^2, m_\rho^2, m_\rho^2] + \lambda_1 \lambda_3 m_\rho^2 \mathrm{C_0} [0,0,0] \notag \\
  &+ 12 \lambda_3^2 m_\rho^2 \mathrm{C_0} [m_{\phi_2}^2, m_\rho^2, m_\rho^2 ] +3\lambda_3^2 m_\rho^2 \mathrm{C_0} [m_\rho^2, m_\rho^2, m_{\phi_2}^2 ] \notag \\
  &+ 4\lambda_2 \lambda_3^2 v^2 \mathrm{C_0} [m_{\phi_2}^2, m_{\phi_2}^2, m_{\phi_2}^2]  + \lambda_3^3 v^2 \mathrm{C_0}[m_{\phi_2}^2, m_{\phi_2}^2,m_\rho^2] \notag \\
    &+4\lambda_3^3 v^2 \mathrm{C_0}[m_{\phi_2}^2, m_\rho^2,m_{\phi_2}^2] + 18 \lambda_3^2 m_\rho^4 \mathrm{D_0} [m_{\phi_2}^2, m_\rho^2, m_\rho^2, m_\rho^2] \notag \\
    &+ 6\lambda_3^3 m_\rho^2 v^2 \mathrm{D_0} [m_{\phi_2}^2, m_\rho^2, m_{\phi_2}^2, m_\rho^2] \notag \\
    &+ 2\lambda_3^4 v^4 \mathrm{D_0}[m_{\phi_2}^2, m_{\phi_2}^2, m_{\phi_2}^2, m_\rho^2],
\end{align}
and
\begin{align}
  &16 \pi^2 \Gamma^{(4)}_{\phi_2 \phi_2^\dagger \phi_2 \phi_2^\dagger} \Big|_{k=\mathrm{sym.}} = -32 \pi^2 Z_{\lambda_2} \lambda_2 +10 \lambda_2^2 \mathrm{B_0} [m_{\phi_2}^2, m_{\phi_2}^2] \notag \\
  &+ \lambda_3^2 \mathrm{B_0} [m_\rho^2, m_\rho^2] + \lambda_3^2 \mathrm{B_0}[0,0] + 12 \lambda_2 \lambda_3^2 v^2 \mathrm{C_0}[m_{\phi_2}^2, m_{\phi_2}^2, m_\rho^2] \notag \\
  &+ 4 \lambda_3^3 v^2 \mathrm{C_0}[m_{\rho}^2,m_{\rho}^2,m_{\phi_2}^2] + 4 \lambda_3^4 v^4 \mathrm{D_0}[m_{\phi_2}^2,m_{\phi_2}^2,m_\rho^2,m_\rho^2].
\end{align}
Here we have defined shorthand notations as
\begin{align}
  &\mathrm{B_0} [a,b] \equiv \mathrm{B_0} \Big( -\frac{4}{3}Q^2; a, b \Big), \notag \\
  &\mathrm{C_0} [a,b,c] \equiv \mathrm{C_0} \Big( -\frac{4}{3}Q^2, -Q^2, -Q^2; a, b, c \Big), \notag \\
  &\mathrm{D_0}[a,b, c, d] \equiv \notag \\
  &\mathrm{D_0} \Big( -\frac{4}{3}Q^2, -Q^2,-\frac{4}{3}Q^2, -Q^2, -Q^2, -Q^2; a, b, c, d \Big).
\end{align}

After imposing the seven renormalization conditions given in eqs.~(\ref{eq:toy_station}), (\ref{eq:toy_rencon1}) and (\ref{eq:toy_rencon2}), the MD beta functions are obtained by 
\begin{align}
  &\beta_{\lambda_1} = \mathcal{D}_Q \lambda_1 = \lambda_1 \mathcal{D}_Q \log(Z_{\lambda_1}^{-1} Z_{\phi_1}^2), \notag \\
  &\beta_{\lambda_2} = \mathcal{D}_Q \lambda_2 = \lambda_2 \mathcal{D}_Q \log(Z_{\lambda_2}^{-1} Z_{\phi_2}^2), \notag \\
  &\beta_{\lambda_3} = \mathcal{D}_Q \lambda_3 = \lambda_3 \mathcal{D}_Q \log(Z_{\lambda_3}^{-1} Z_{\phi_1} Z_{\phi_2}).
\end{align}

\bibliography{references}

\end{document}